\documentclass[notitlepage,a4paper,12pt]{article}
\usepackage{sw20elba}
\usepackage[dvips]{graphicx}

\begin{document}

\author{\textbf{George Jaroszkiewicz} \\
%EndAName
School of Mathematical Sciences, University of Nottingham, UK}
\title{\textbf{Discrete spacetime: classical causality,}\\
\textbf{prediction, retrodiction and the}\\
\textbf{mathematical arrow of time}\thanks{%
Talk given at The First International Interdisciplinary Workshop on
''Studies on the Structure of Time: from Physics to Psycho(patho)logy'',
23-24 November 1999, Consiglio Nazionale delle Richerche, Area della Ricerca
di Palermo, Palermo, Sicily, Italy.}} 
\date{}
\maketitle

\begin{abstract}
A mathematical definition of classical causality over discrete spacetime
dynamics is formulated. The approach is background free and permits a
definition of causality in a precise way whenever the spacetime dynamics
permits. It gives a natural meaning to the concepts of cosmic time,
spacelike hypersurfaces and timelike or lightlike flows without assuming the
notion of a background metric. The concepts of causal propagators and the
speed of causality are discussed. In this approach the concepts of spacetime
and dynamics are linked in an essential and inseparable whole, with no
meaning to either on its own.
\end{abstract}

\section{\protect\smallskip \textbf{Introduction}}

The term \textit{causality} is frequently used in a way which suggests that
an intrinsic causal structure underlies the universe. In relativity this is
reinforced by the assumption of a metric tensor with a Lorentzian signature.
This gives the traditional light cone structure associated with spacelike
and timelike intervals, and imposes conditions on the possible trajectories
of particles and quantum field theory operator commutation relations.

We shall discuss the idea that \textit{causality is a convenient account
designed to satisfy and conform to the patterns of classical logic that the
human Theorist wishes to believe underlies the dynamics of space, time and
matter. }In this approach causality need not be associated with any \textit{%
a priori }concept of metric tensor.

This view of causality has been suggested by various philosophers and
scientists. Hume argued that causality is a fiction of the mind. He said
that causal reasoning is due to the mind's expectation that the future is
like the past and because people have always associated an effect with a
cause. Kant believed that causality is a category used to classify
experience. Lentzen said that causality is a relation within the realm of
conceptual objects. Lurchin said that causality is a personal way of thought
that is not among our immediate sensual data, but is rather our basic way to
organize that data. For Maxwell the principle of causality expresses the
general objective of theoretical sciences to achieve deterministic
explanations$,$ and according to Heisenberg, causality only applies to
mathematical representations of reality and not to individual mechanical
systems.

\section{\textbf{The PPM view of time}}

To avoid confusion we distinguish three sorts of time:

$i)\ $\ \ \textit{Process Time} is the hypothesized time of physical
reality. Although there is geological and astrophysical evidence for some
sort of temporal ordering in reality \cite{WHITROW:80}, process time need
not exist in any real sense and may just be a convenient way for humans to
think about the Universe.

$ii)$\ \ \textit{Physiotime} is the subjective time that humans sense and
which they believe runs monotonically forwards for them. It is the end
product of complex bio-dynamical processes occurring in process time. Its
origins are not understood currently. Many physicists believe that this
feeling is an illusion. What matters here is the undeniable \textit{existence%
} of this feeling, because humans are driven by this sensation of an ever
increasing time to believe that descriptions of reality must involve such a
concept.

$iii)$\ \textit{Mathematical times }are conceptual inventions of human
theorists designed to model process time. Examples are Newtonian Absolute
Time, relativistic coordinate times, proper time and cosmic time.
Mathematical times usually involve subsets of the real line, which has an
ordering property. This ordering is used to model the notions of \textit{%
earlier} and \textit{later.} This presupposes something about the nature of
process time that may be unwarranted. In the Euclidean formulation of field
theory for example there is no dynamical ordering parameter.

\subsection{\textbf{The Theorist}}

We shall use the term \textit{Theorist} to denote the human mind operating
at its clearest and most rational in physiotime. The Theorist has the status
of an observer or deity overseeing the mathematically consistent development
of chosen mathematical models that are used to represent phenomena in
process time. \textit{Free will} enters into the discussion here as the
freedom of the Theorist to choose boundary conditions in these models.
Whether free will is an illusion or not is regarded here as irrelevant.

\section{\textbf{Classical causality}}

The need to seek causal explanations stems from the peculiarities of human
consciousness. Humans generally want to \textit{explain }phenomena. When
they do this they invariably try to invoke what may be called \textit{%
classical logic. }This is the everyday logic that postulates that statements
are \textit{either true or else false} and that conclusions can be drawn
from given premises. It is also the logic of vision, which generally informs
the brain that an object either is in a place or else is not in that place.
The rational conscious mind tends to believe that the external universe
follows this logic, and this is the basis for the construction of CM
(classical mechanics) and all the belief structures which it encodes into
its view of reality. It is also the logic of jurisprudence and common sense.
This logic served humanity extremely well for millennia, until technological
advances in the early years of the twentieth century revealed that quantum
phenomena did not obey this logic in detail.

A CM Theorist is anyone who believes in a classical view of reality. In the
mindset of a CM Theorist, reality is assumed to be strictly single valued at
each and every time even in the absence of observation. Philosophers say
that reality is determinate\textit{. }The CM Theorist attempts to make
unique predictions wherever possible, such as where a planet will be at a
future time. The assumption is made that the planet will be somewhere at
that time and not nowhere, and that it will not be in two or more places at
that time.

\smallskip In general, quantum theory requires a pre-existing classical
conceptual framework for a sensible interpretation. For example,
relativistic quantum field theory assumes a classical Lorentzian metric over
spacetime, and only the fields are quantised. For this reason, we shall
focus our attention on a classical formulation of causality.

\section{\textbf{Functions and links}}

To set up our framework for causality, it will be useful to review the
definition of a \textit{function}:

\subparagraph{\textbf{Definition 1:}}

A\textit{\ function} $f$ \cite{HOWSON:72} is an ordered triple $f\equiv
(F,D,R)$ where $F,\,D$ and $R$ are sets which satisfy the following:

\begin{enumerate}
\item  $F$ is a subset of the Cartesian product $D\times R$;

\item  for each element $x$ in $D$ there is exactly one element $y$ in $R$
such that the ordered pair $\left( x,y\right) $ is an element of $F$.
\end{enumerate}

Physicists tend to write $y=f\left( x\right)$ . $D$ is called the \textit{%
domain} of (definition of) $f$ and $R$ is the \textit{%
range\ }of\textit{\ }$f.\;$ The \textit{image\ }of\textit{\ }$f$ is the
subset $f(D)$ of $R$ such that for each element $v$ in $%
f( D) $ there is at least one element $u$ in $D$ such that $%
v=f( u) .$

Without further information it cannot be assumed that $f\left( D\right) =R,$
but in our work this must be assumed to hold. Otherwise, there arises the
possibility of having a CM where something could happen without a cause,
i.e. an element $z$ of $R$ could exist for which there is no $x$ in $D$ such
that $z=f\left( x\right) $.

The range and domain of a function do not have a symmetrical relationship
and the ordering of the sets in \textit{Definition }$1$ is crucial. Usually,
no pair of component sets in the definition of a function can be
interchanged without changing the function. This asymmetry forms the basis
of the time concept discussed in this article and defines what we call a 
\textit{mathematical arrow of time}. Assuming that $f$ is single valued,
then we may employ the language of dynamics here, though this may seem
unusual. We could say that $y=f(x)$ is \textit{determined} by $x$\textit{\
via the process }$f$\textit{, }or that $x$\textit{\ causes }$f\left(
x\right) $. Then $x$ is a \textit{cause}, $f\left( x\right) $ is its \textit{%
effect }and $f$ is the \textit{mechanism of causation}.$\;\mathcal{\,}$

Although \textit{Definition} $1$ carefully excludes the concept of a
many-valued explicit function, such a possibility arises when we discuss
implicit functions.

\subparagraph{Graphical notation:}

We shall use the following graphical notation:

\subsection{\textbf{Explicit functions:}}

The process of mapping elements of $D$ into $R$ via $f\;$ will be
represented by the LHS (left hand side) of $Fig.1a$, where the large circles
denote domain and image sets, the small circle denotes the function, and
arrows indicate the direction of the mapping. An alternative representation
is given by the RHS (right hand side) of $Fig.1a$, where the labels $D$, $R$
and $f$ are understood. Here the integers $0$ and $1$ represent the ordering
of the function \textit{from }$D$\textit{\ }$\left( \equiv 0\right) $ 
\textit{to }$R\left( \equiv 1\right) $, so that the arrows are not needed.

\begin{figure}
\begin{center}
\includegraphics[width=3.97in,height=2.57in]{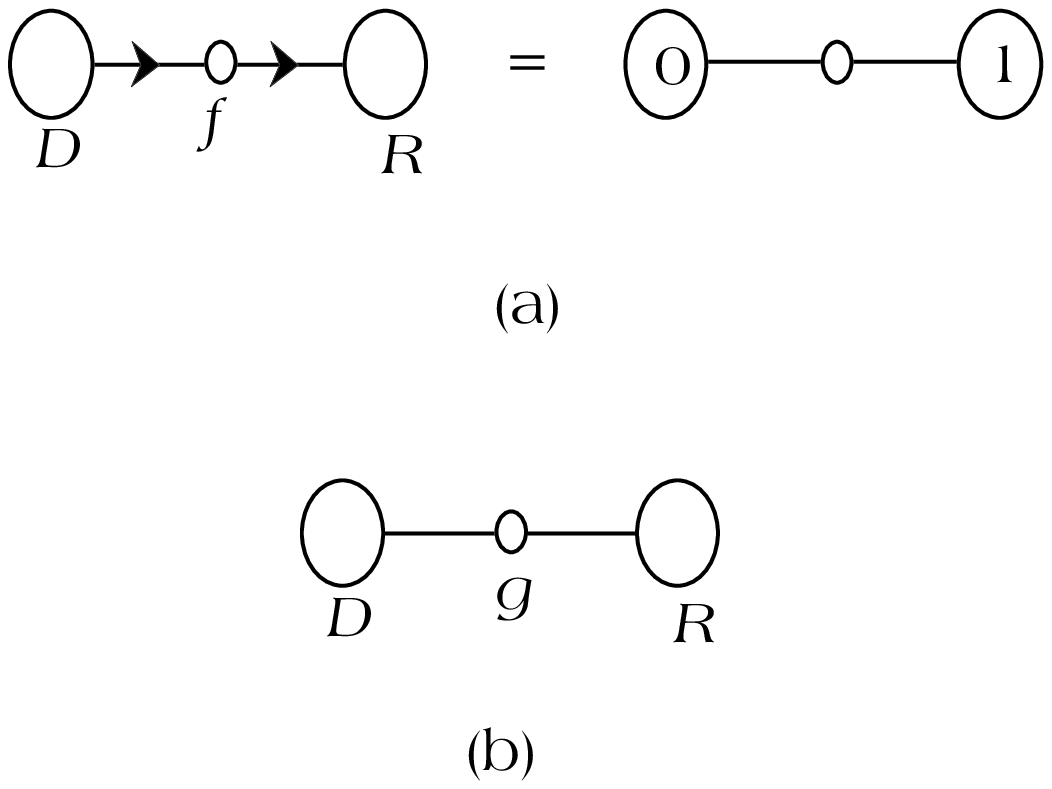}
\caption{(a) The LHS represents
a function $f$ with domain $D$ and range $R$; the RHS is an equivalent
representation of the same with $D$, $R$ and $f$ understood. The ordering of
the digits $0$, $1$ represents the direction of the mapping. (b) A
representation of an implicit function $g$ of two variables.}
\end{center}
\end{figure}

\subsection{\textbf{Many-to-one functions:}}

\textit{Definition} $1$ raises the question: \textit{which particular} $x$ 
\textit{in} $D$ \textit{caused a given} $y$ \textit{in} $R?$ As given,
nothing in \textit{Definition} $1$ rules out many-to-one mappings, so
without further information about the function $f,$ there may be more than
one such $x.$ The assumption of a unique pre-image is equivalent to the
belief that, given the present state of the universe, there was a unique
past which gave rise to it. 't Hooft has recently discussed this in the
context of gravitation using equivalence classes of causes \cite{T'HOOFT-99}%
. This is bound up with the notion of irreversibility.

If however it is the case that $f$ is one-to-one and onto, then its inverse
function $f^{-1}$ exists and so any element $y=f\left( x\right) $ in $R$ can
be mapped back to a unique cause $x$ in $D$ via $f^{-1}.\;$In such a case
there would be no inherent difference in principle between the roles of $D$
and $R$.\ In the language of dynamics the mechanics would be \textit{%
reversible.\ }

\subsection{\textbf{Implicit functions and links:}}

Suppose now that the relationship between $D$ and $R$ is implicit rather
than explicit. For example, let $D$ and $R$ each be a copy of the real line\ 
$\mathsf{R}$ with elements $x$,$\;y$ belonging to $D$ and $R$ respectively
and suppose that the only dynamical information given was an implicit
equation of the form 
\begin{equation}
g\left( x,y\right) =0.  \label{aaa}
\end{equation}
Then our graphical notation for this is given by $Fig.1b$, where the small
circle now denotes an implicit equation or \textit{link} $g$ relating
elements of $D$ and $R$.\ Without further information no arrows are
permitted at this stage.

\subsection{\textbf{Resolution of links:}}

Given an implicit equation in two variables such as $\left( \ref{aaa}\right) 
$, suppose now that it could proved that there was always a unique solution
for $y$ in $R$ given any $x$ in $D$ (the solution $y$ of course depending on
the value of $x)$. Then our convention is that now arrows pointing from $D$
into $R$ via the link $g$ may be added to indicate this possibility, giving
the LHS of $Fig.1a$ with $f$ replaced by $g$. But now the relationship
between $D$ and $R$ is formally equivalent in principle to having an
explicit function $y$ of $x$ (even if $y$ could not be obtained
analytically). In such a case we will also say that any given element of $D$ 
\textit{causes} or \textit{determines }a corresponding element of $R$ and
that the link between $D$ and $R$ may be \textit{resolved} from $D$ \textit{%
to} (or in favour of) $R$.

Our concept of resolution depends only on the \textit{existence} of a unique
solution and does not imply that a solution could actually be computed by
the Theorist in practice. Computability is an attribute associated with 
\textit{physiotime}, and is not here regarded as an essential ingredient of
our version of causality. A more severe definition of causality however
might be to impose the restriction of computability. We do not do this here
because we wish to avoid anthropomorphism. What is important in classical
mechanics is the existence of a unique resolution; the universe does not
actually ``compute'' anything when this resolution occurs.

\subsection{\textbf{Inequalities:}}

It is possible to consider links which are not equations but more general
relations such as inequalities. For example, suppose $D$ and $R$ are copies
of the real line with elements denoted by $x$, $y$ respectively and consider
a link defined by 
\begin{equation}
g(x,y)\equiv x+y<0.
\end{equation}
Given $x$ there is an infinity of solutions for $y$, so in this case one
possible interpretation would be that the link is equivalent to a many
valued function of $x$, although this way of putting things might seem
unusual. On the other hand, given a $y$ there is also an infinity of
solutions for $x$. This leads to an alternative interpretation of the link
as a many valued function of $y$.

Such examples do not generate a classical TRA (temporal resolution of
alternatives) and so would not occur normally in our classical spacetime
dynamics.

\subsection{\textbf{Reversibility:}}

Now suppose we were given an implicit equation for $x$ and $y$ such that we
could prove the existence of a unique solution for either $x$ or $y$ given
the other variable. Then the arrows could point either way and this would
correspond to a choice of causation\footnote{%
This choice is taken by the Theorist in physiotime.}. Because cause and
effect can be interchanged in such a case, it then becomes meaningful to
talk about this dynamics being \textit{reversible. }Clearly this is formally
equivalent to having an invertible explicit function.

When discussed in this way it becomes clear that in general, spacetime
dynamics will be irreversible. Reversibility will occur only under very
special conditions, which of course is the experience of experimentalists.

\subsection{\textbf{Generalization:}}

In general the sets $D$ and $R$ need not be restricted to the reals. They
could be any sort of sets, such as vector spaces, tensor product spaces,
quaternions, operator rings, group manifolds, etc. Whatever their nature,
they will always be referred to as \textit{events} for convenience. In the
sorts of dynamics we have in mind events will often be sets such as group
manifolds.

\textit{Links} are defined as specific relationships between events. They
may be more complicated than those in the above examples, and may have
several or many components relating different components of different
events. The specification of a set of events and corresponding links will be
said to specify a (classical) \textit{spacetime dynamics.\ }

An important generalization is that a link might involve more than two
events. Given for example five events $R$, $S$, $T$,\ $\;U$ and $V$ with
elements $r,\;s,\;t,\;u$ and $v$ respectively then a classical dynamics
involving these events would give some set of equations or link $g$ of the
generic form 
\begin{equation}
g\left( r,s,t,u,v\right) =0.
\end{equation}
This will be represented by $Fig.2a$.

\begin{figure}[!h]
\begin{center}
\includegraphics[width=4.4in,height=5.33in]{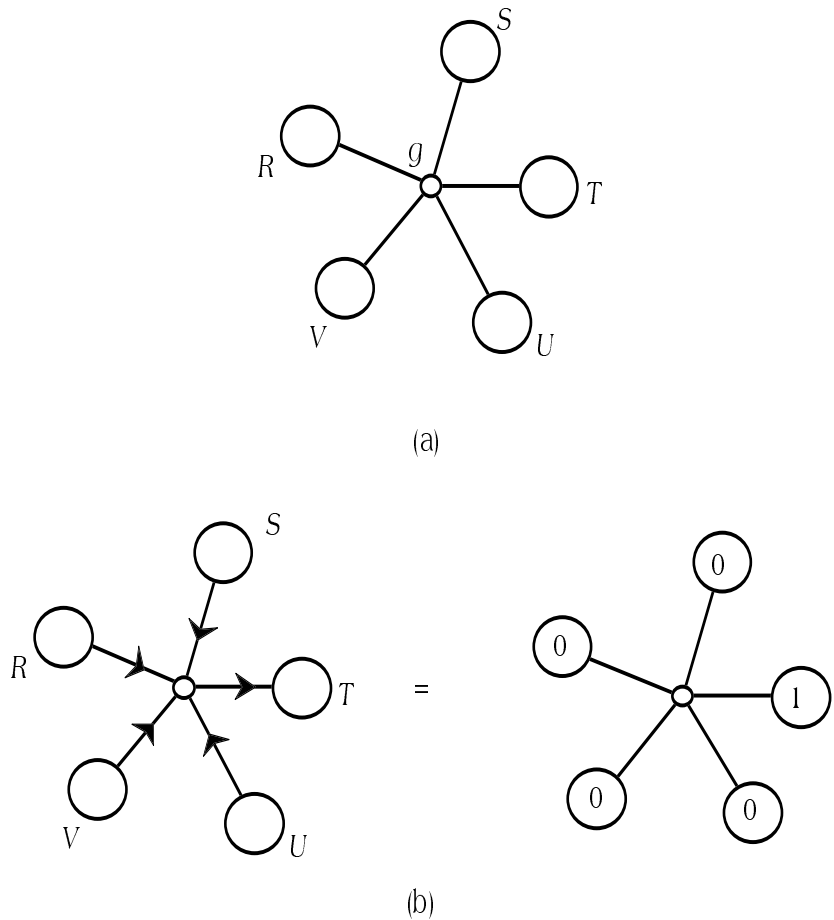}
\caption{(a) An implicit
equation or link involving elements from five events. (b) If the link can be
resolved in favour of $T$ then arrows are added as on the LHS. An equivalent
diagram is given on the RHS with the ordering of the integers indicating the
direction of the resolution.}
\end{center}
\end{figure}

Suppose now, given such an $g$, that it could be proven that there is always
a unique solution $t\in T$ given the other elements $r,$ $s,$ $u$ and $v$.
In such a case this will be indicated by arrows pointing from $R$, $S$, $U$
and $V$ into $g$ and an arrow pointing into $T$ as in the LHS of $Fig.2b$.
Then it will be said that $g$ \textit{can be (causally) resolved in} \textit{%
(favour of)} $T$, and $T$ will be called the \textit{resolved event}.

By definition, classical resolution is always in favour of a single resolved
event, given initial data about the other events associated with the link.
This does not imply anything about the possibility of resolving $g$\textit{\ 
}in favour of any of these other events. It may be or not be possible to do
this.

Suppose the Theorist could in principle resolve $T$ if they were given $%
R,S,U $ and $V$, and also resolve $S$ if they were given $R,T,U$ and $V$.
Then the Theorist has to make a choice of resolution and choose one possible
resolution and exclude the other. It would not be meaningful classically to
resolve $g$ in favour of $T$ and $S$ at the same moment of physiotime, the
reason being that these alternative resolutions employ different and
inconsistent initial data sets (boundary conditions). Initial data sets are
equivalent to information, and it is a self-evident premise that a Theorist
can have at most one initial data set from a collection of possible and
mutually inconsistent initial data sets at a given moment of physiotime.

There remains one additional exotic possibility. If might be the case that
given say $R,S$ and $T$, both $U$ and $V$ were implied by a knowledge of the
link. For example, suppose the link was equivalent to the equation 
\begin{equation}
r+s+it+u+iv=0,
\end{equation}
where $i=\sqrt{-1}$. Assuming that $r,s,t,u$ and $v$ were always required to
be real, then we could always find $u$ and $v$ from a given $r,s$ and $t,$
simply by equating real and imaginary parts. A situation where a given link
can be resolved in two or more events given just one initial data set at
that link will be called a \textit{fluctuation process. }Fluctuation
processes will be excluded from our notion of classical causality. They may
have a role in the QM (quantum mechanics) version of causality, which is
beyond the scope of this article.

One reason for excluding fluctuation processes is that this guarantees that
information flows (in physiotime) from a link into a \textit{single} event.
This is related to the concept of \textit{cosmic time} discussed below and
to the idea that classical mathematical time has one dimension. When
theorists discuss models with more than one parameter called a time, all but
one of these has to be hidden or eliminated at the end of the day if a
classical picture is to emerge.

It may be argued therefore that \textit{the mechanism of classical
resolution is the origin of the concept of time, and that time, like
causality, is a no more than a convenient theoretical construct designed by
the human mind to provide a coherent description of physical reality.} This
carries no implication that what we called process time is really a linear
time. As we said before, process time is just a convenient label for
something which may be quite different to what we believe it to be.

When a choice of resolution exists and is made, then as an additional
simplification and provided there is no confusion with other relations to
which $T$ may be a party (not shown), the diagram on the LHS of $Fig.2b$ may
be replaced by the RHS of $Fig.2b$. Here the numbers zero and one indicate
the ordering of the resolution. Because $T$ may be regarded as \textit{caused%
} by the other events, it can be regarded as later and so has a greater
associated discrete time. Such times will be called \textit{dates}.

In general a link may be a whole collection of relations and equations. If
there is just one small part of these equations which does not determine a
unique solution fully, i.e., prevents a resolution, then arrows or dates are
in principle not permitted. However, under some circumstances it may be
reasonable to ignore some part of a dynamical relation in such a way that
arrows could be justified as far as the remaining parts were concerned. For
example, the microscopic laws of mechanics appear to be resolvable forwards
and backwards in time (i.e., are reversible) provided the ``small'' matter
of neutral kaon decays and the thermodynamic arrow of time (which could
involve the gravitational field \cite{KAY-98}) are ignored.

In $Fig.1a$, $D$ is the \textit{complete cause} of $R$; in $Fig.2b$, $R$ is
a \textit{partial cause} of $T$. The \textit{complete cause} of $T$ is the
collection of events $R$, $S$, $U$ and $V$, but only for this choice of
resolution. The Theorist could decide to alter boundary conditions so that $%
T $ was no longer regarded as the resolved event.

Having outlined our ideas on functions and links, we shall apply them now to
discrete spacetime.

\section{\textbf{Discrete spacetime}}

Fourier's \textit{principle of similitude} states that a system $S^{\prime }$
similar to but smaller than another system $S$ should behave like $S.$ It is
the physicist's analogue of continuity in mathematics, and is of course an
erroneous principle when applied to matter, as evidenced by the observation
of atoms and molecules. It is generally supposed that this principle will
also break down in the microscopic description of space and time. Classical
GR (general relativity) may therefore be an approximation, albeit a
remarkably good approximation, to some model of space and time which is not
intrinsically a four-dimensional pseudo-Riemannian manifold.

There have been numerous suggestions concerning the fundamental nature and
meaning of space and time, such as twistor theory, point set theory, etc., 
\cite{ISHAM-99} and each of these suggestions makes a specific set of
mathematical assumptions about spacetime. Likewise, in this article a
specific view of space time and dynamics is proposed and its consequences
explored. Of course, there is an important question concerning the use of
classical or quantum physics here. In this paper the proposals are based on
classical ideas and the ramifications of quantum physics are explored
elsewhere.

From before the time of Newton, physicists took the view that material
objects have definite spatial positions at definite times. In the $20^{th}\;$%
Century theorists went further and developed to the extreme the Wellsian or
block Universe \cite{PRICE:97} view that space and time exist in some
physically meaningful sense, even in the absence of matter. Whatever that
sense is, be it a physical one or simply an approximate relationship of
sorts between more complex attributes of reality, most physicists agree on
the prime status of spacetime as the arena in which or over which physical
objects exist. This is certainly the case for classical physicists and to
various degrees for quantum physicists.

In this paper the focus is on a classical description of a \textit{discrete}
spacetime structure. Discreteness is considered here for several reasons.
First, as mentioned above, it would be too much to hope that Fourier's
principle of similitude should apply to spacetime and not to matter. Second,
there is a strong feeling in the subject of quantum gravity that the Planck
scales are significant. Third, discreteness has the advantage over
continuity in being less mathematically restrictive. Theories based on
discrete principles can usually encompass the properties of those based on
continuous principles via appropriate limit processes, yet retain features
which cannot occur in the continuum.

Discrete spacetime structure and its relationship to causality has been
discussed by a number of authors, notably by Sorkin et al \cite{SORKIN+al-87}
and Finkelstein et al \cite{FINKELSTEIN-74B}. A basic difference between
those approaches and that taken here is that no \textit{a priori }underlying
spacetime manifold is assumed here.

\subparagraph{\textbf{Proposition }$1$:}

classical spacetime may be modelled by some discrete set $\mathcal{S}$.

Elements of $\mathcal{S}$ will be denoted by capital letters such as $%
P,\,Q,\,R,\;$etc. $\mathcal{S}$ will be called a \emph{spacetime} and its
elements referred to as \emph{events} even if subsequently $\mathcal{S}$
turns out to have only an indirect relationship with the usual
four-dimensional spacetime of physics. What these events mean physically
depends on the model. It is simplest to think of events as labels for
mathematical structures representing the deeper physical reality associated
with process time.

Events are meaningful only in relationship with each other and it is
meaningless to talk about a single event in spacetime without a discussion
of how the Theorist relates it to other events in spacetime. This is done by
specifying the \textit{links }or relationships between the events. Links are
as important as the events themselves and it is the totality of links and
events which makes up our spacetime dynamics. This should include all
attributes relating to matter and gravitation, and it is in principle not
possible to discuss one without the other.

The structure of our spacetime dynamics is really all there is; links and
events. No preordained notion of metrical causality involving spacelike and
timelike intervals is assumed from the outset. All of that should emerge as
part of the implications of the theory. In classical continuous spacetime
theories, on the other hand, the metric is usually assumed to exist
independently of any matter, even before it is found via the equations of
GR. As we said before, this metric carries with it lightcone structure and
other pre-ordained attributes of causality.

Discrete spacetime carries with it the astonishing possibility of providing
a natural explanation for length, area and volume. According to this idea,
attributed to Riemann \cite{SORKIN+al-87}, these are simply \textit{%
numerical counts} of how many events lie in certain subsets of spacetime.
Discreteness may also provide a natural scale for the elimination of the
divergences of field theory, and permits all sorts of novelties to occur
which are difficult if not impossible to build into a manifold. Recently,
the study of spin networks in quantum gravity has revealed that quantization
of length, area and volume can occur \cite{ROVELLI-98}.

\section{\textbf{Event state space}}

In CM the aim is usually to describe the temporal evolution of chosen
dynamical degrees of freedom. These take on many possible forms, such as
position coordinates or various fields variables such as scalar, vector and
spinor fields. In our approach, we associate with each event $P$ in a given
spacetime $\mathcal{S}$ an internal space $\Omega _P$ of dynamical degrees
of freedom called \textit{event state space}. This space could be whatever
the Theorist requires to model the situation. Moreover, it could be
different in nature at each event $P$ in this spacetime. Elements of $\Omega
_P$ will be denoted by lower case Greek letters, such as $\xi _p,\lambda _P$
etc. and a chosen element $\xi _P$ of $\Omega _P$ will be called a \textit{%
state} \textit{of the event} $P$.

Elementary examples of event state spaces are:

\subparagraph{$i)$ \textbf{Scalar fields}:}

a real valued scalar field $f$ on a spacetime $\mathcal{S}$ is simply a rule
which assigns at each event $P\in \mathcal{S}$ some real number $f_p.$ This
may be readily generalized to complex valued functions. If no other
structure is involved then obviously 
\begin{eqnarray}
{\Omega}_P &=&\mathsf{R}_P\;\;\;\;\forall P\in \mathcal{S}, \nonumber \\
{\xi}_P &\equiv &f_P\in \mathsf{R}_P,
\end{eqnarray}
where $\mathsf{R}_P$ is a copy at $P$ of the real line $\mathsf{R.}$

A classical configuration of spacetime in this model would then be some set $%
\left\{ f_P,\;f_Q,...\right\} $. This configuration ``exists'' at some
moment of the Theorist's physiotime but the model itself would not
necessarily have any causal ordering. That would have to be determined by
the Theorist in the manner discussed below.

\subparagraph{$ii)$ \textbf{Vector fields: }}

suppose at each event $P$ there is a copy $V_P$ of some finite dimensional
vector space $V\;$with elements $\mathbf{v}\in V,$ etc. Then a vector field
is simply a rule which assigns at each event $P$ some element $\mathbf{v}%
_P\in V_P.$

\subparagraph{$iii)$ \textbf{Group manifolds:}}

at each event P we choose a copy $G_P$ of some chosen abstract group, such
as $Z_2$ or $SU(3)$ to be our event space.

\subparagraph{$iv)$ \textbf{Spin networks: }}

A spin network is a graph with edges labelled by representations of a Lie
group and vertices labelled by intertwining operators. Spin networks were
originally invented by Penrose in an attempt to formulate spacetime physics
in terms of combinatorial techniques \cite{ROVELLI+SMOLIN-95} but they may
also defined as graphs embedded in a pre-existing manifold \cite{BAEZ-95}.
The state event space at each event $P$ is a copy of $SU(2).$ In this
particular model the sort of events we are thinking of in spacetime would be
associated with the geometrical links of a triangulation and the links (the
dynamical relationship between our events) would be associated with the
geometrical vertices of the triangulation, that is, with the intertwining
operators.

\subsection{\textbf{Neighbourhoods and local environments}}

For physically realistic models the number of events in the corresponding
spacetime $\mathcal{S}$ will be vast, possibly infinite. Sorkin et al \cite
{SORKIN+al-87} give a figure of the order $10^{139}$ per cubic
centimetre-second, assuming Planck scales for the discrete spacetime
structure. We shall find it useful to discuss some examples with a finite
number of events for illustrative purposes.

In our spacetime diagrams we will follow the convention established for
functions in the previous section; large circles denote events and small
circles denote links, with lines connecting events and links. Before any
temporal resolution is attempted, no arrows can be drawn. $Fig.3a$ shows a
finite spacetime with 14 events and 9 links.

\subparagraph{\textbf{Definition }$2$\textbf{: }}

The ( \textit{local})  \textit{environment }$\mathcal{E}%
_A$ \textit{of an event} $A$ is the subset of links which involves $A$, that
is, all those links to which $A$ is party, and the \textit{degree of an event%
} is the number of elements in its local environment.

For example, from $Fig.3a$, the local environment of the event labelled $P$
is the set of links\ $\mathcal{E}_P\equiv \left\{ f,g,h\right\} $, and so $P$
is a third degree event.

\subparagraph{\textbf{Definition }$3$\textbf{: }}

The \textit{neighbourhood }$\mathcal{N}_A$ \textit{of an event }$A$ is the
set of events linked to $A$ via its environment.

For example, from $Fig.3a$\ the neighbourhood of event $P$ is the set of
events $\mathcal{N}_P\equiv \left\{ Q,R,S,T,U,V\right\} ,$ and $Q,R,S,T,U$
and $V$ are the \textit{neighbours} of $P.$

\subparagraph{\textbf{Definition }$4$\textbf{: }}

The \textit{domain }$\mathcal{D}_f$ \textit{of a link }$f$ is the set of
events involved in that link, and the \textit{order of a link }is the number
of elements in its domain.

For example, from $Fig.3a$, $\mathcal{D}_f$ $\equiv \left\{ Q,R,P,U\right\} $
and so $f$ is a fourth order link.

The local environment of an event will be determined by the underlying
dynamics of the spacetime, i.e. the assumed fundamental laws of physics.
Currently these laws are still being formulated and discussed, so only a
more general (and hence vague) discussion can be given here with some simple
examples. A spacetime and its associated structure of neighbourhoods and
local environments will be called a \textit{spacetime dynamics.}

\begin{figure}[!h]
\begin{center}
\includegraphics[width=4.74in,height=6.77in]{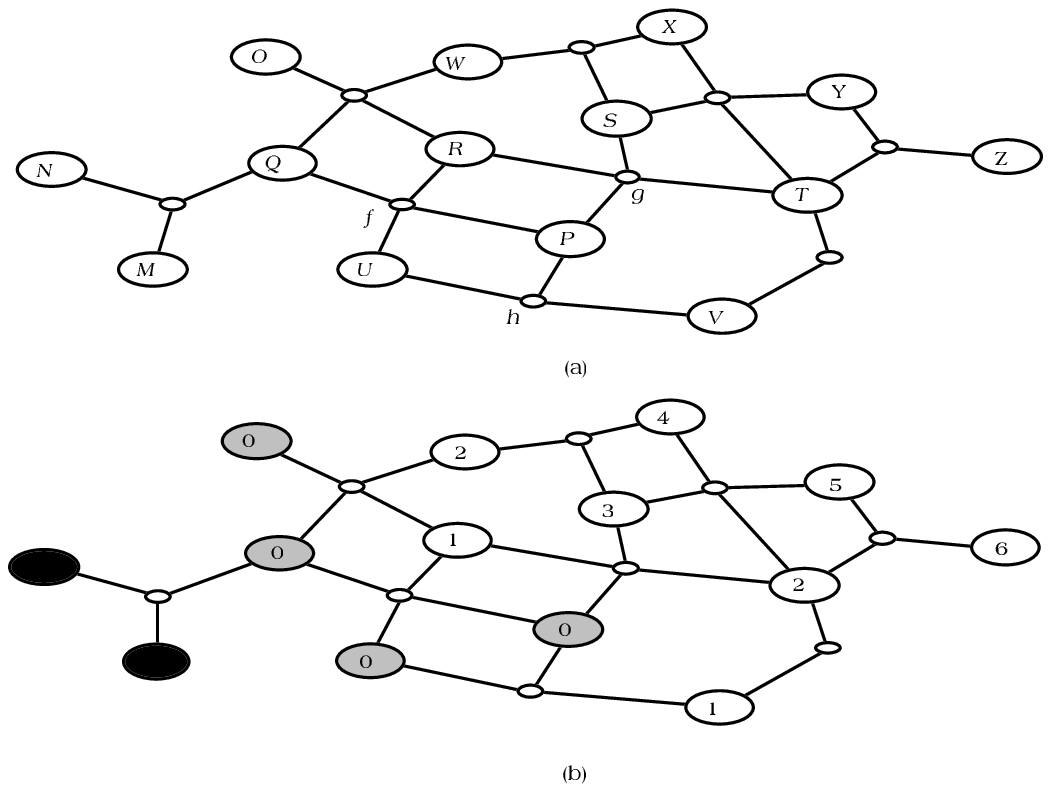}
\caption{(a) $\{f,g,h\}$ is the
local environment of $P$ and $\{Q,R,S,T,U,V\}$ is its neighbourhood. $%
\{Q,P,R,U\}$ is the domain of $f$. (b) A chosen initial data set $\{O,P,Q,U\}
$ is shaded grey. Its full implication (or future) is $\{R,S,T,V,W,X,Y,Z\}$.
Its (absolute) past $\{M,N\}$ is inaccessible from this initial data set and
cannot be influenced by any changes on the initial data set.}
\end{center}
\end{figure}

\section{Kinds of spacetime dynamics}

There are two interrelated aspects of any spacetime dynamics: $i)$ the
nature of the event state space associated with each event and $ii)$ the
nature of the links connecting these events. The way in which events are
related to each other structurally via the links will be called the \textit{%
local discrete topology.} If this discrete topology has a regularity holding
for all links and events, such as that of some regular lattice network, then
we shall call this a \textit{homogeneous} discrete topology. Otherwise it
will be called \textit{inhomogeneous. }

We envisage three classes of classical spacetime dynamics:

\begin{enumerate}
\item  \textbf{Type }$\mathbf{A}$\textbf{:} spacetime dynamics with a
homogeneous and fixed discrete topology with a variable\footnote{%
i.e., variable in physiotime.} event state configuration which does not
affect the discrete topology. This corresponds to (say) field theory over
Minkowski spacetime;

\item  \textbf{Type }$\mathbf{B}$\textbf{:} spacetime dynamics with
inhomogeneous but fixed discrete topology with a variable event state
configuration which does not affect the discrete topology. This corresponds
to field theory over a fixed curved background spacetime, such as in the
vicinity of a black hole;

\item  \textbf{Type }$\mathbf{C}$\textbf{:} spacetime dynamics with discrete
topology determined by the event state configuration; this corresponds to GR
with matter.
\end{enumerate}

Type $A$ and $B$ spacetime dynamics are relatively easy to discuss. Once a
fixed discrete topology is given this provides the template or matrix for
the Theorist to ``slot in'' the causal patterns associated with initial
event state configurations (initial data sets). In this sense, types $A$ and 
$B$ are not genuinely background free, but they are independent in a sense
of any preordained Lorentzian metric structure\textit{.}

Type $C$ presents an altogether more interesting scenario to discuss and
demonstrates the basic issue in classical GR which is that the spacetime
dynamics should determine its own structure, including its topology.

GR without matter of any sort does not make sense in our approach, because
we need to specify event state spaces in order to define the links. Spins
are needed to specify spin networks, for example. In our approach,
gravitation is intimately bound up with discrete spacetime topology.

A spacetime diagram need not be planar. Indeed, there need not be any
concept of spacetime dimension at this stage. Sorkin et al \cite
{SORKIN+al-87}\ suggest that at different scales, a given discrete spacetime
structure might appear to be approximated by different continuous spacetime
dimensions, such as $26$, $10$, or $4,$ depending on the scale.

\section{\textbf{Classical resolution and causal structure}}

The PPM model was introduced as an approach to the modelling of process
time. Working in physiotime, the Theorist first decides on a spacetime
dynamics and from an initial data set then determines a consistent event
state at each event in the spacetime. Because of the existence of the links,
however, these event states cannot all be independent, and this
interdependence induces our notion of causality, as we now explain.

First, restrict the discussion to Type $A$ and $B$ spacetimes and suppose
that each link is \textit{fully resolvable}. By this is meant that if the
order of a link is $n$ and the event states are specified for any $n-1$ of
the events in the domain of the link, then the remaining event space in the
domain is uniquely resolved.

\subparagraph{\textbf{Example }$1$\textbf{:\ \protect\smallskip }}

An example of a fully resolvable link is the following: Let $f$ be a link of
order $r$, with domain $\mathcal{D}_f\equiv \left\{ P_1,P_2,...,P_r\right\}
.\;$Suppose the event space at $P_i$ is a copy $G_i$ of some group $G$ such
as $Z_2$ or $SU(n)$ and suppose the link is defined by 
\begin{equation}
f:g_1g_2...g_r=e,
\end{equation}
where $g_i\in G_i$ and $e$ is the group identity\footnote{%
We may use a matrix representation to define products of elements from
different copies of $G.$}. Then clearly, 
\begin{equation}
g_1=g_r^{-1}g_{r-1}^{-1}...g_2^{-1}
\end{equation}
and similarly for any of the other events states. In other words, we can
always resolve any one of the events in $\mathcal{D}_f$ uniquely in terms of
the others.

Now suppose the Theorist chooses one link, such as $f$ in $Fig.3a$, which is
a fourth order link. Then its domain $\mathcal{D}_f$ can be identified
immediately from the spacetime dynamics to be $\left\{ P,Q,R,U\right\} $.
The Theorist is free to specify the states at three of these without any
constraints, and this represents an initial data set called $\mathcal{S}_0$.
Suppose these are events $P,Q$ and $U$.

If now the structure of the link $f$ is such that there is only one possible
solution in $\Omega \left( R\right) $, then we have a classical resolution
of alternatives\footnote{%
i.e., the alternatives which form the event space $\Omega \left( R\right) .$}
and the emergence of a causal structure. We can use the language of dynamics
and say that events $P,Q$ and $U$ \textit{cause} $R$ and denote it by a
diagram such as $Fig.2b$. But it should be kept in mind that the Theorist
has \textit{decided on which three sets to use as an initial data set. }Our
interpretation of causality is that it is dictated partly by the spacetime
dynamics and partly by the choices made by the Theorist. In general,
classical resolution involves using information about $n-1$ event states at
an $n^{th}$ order link to determine the state of the remaining event in the
domain of that link.

An initial data set may involve more than one link, such as shown in $Fig.3b$%
. In that diagram, events are labelled by integers representing the discrete
times at which their states may be fixed. Events in the chosen initial data
set are labelled with a time $0$ and shaded grey. These are the events $%
\left\{ O,P,Q,U\right\} .$

Given an initial data set $\mathcal{S}_0$, the Theorist can then use the
links to deduce the \textit{first (or primary) implication }$\mathcal{S}_1$.
This is the set labelled by integer time $1$ in $Fig.3b$, and consists of
events $\left\{ R,V\right\} .$ The event state at each of the events in the
primary implication is determined from a knowledge of the event states on
the initial data set, assuming the links do indeed permit a classical TRA.
It could be the case that the primary implication is the empty set.

Given a knowledge of the event states on $\mathcal{S}_0$ and its primary
implication $\mathcal{S}_1,$ then the \textit{second }(or \textit{secondary})%
\textit{\ implication }$\mathcal{S}_2\equiv \left\{ W,T\right\} $ can now be
found and its events labelled by the discrete time $2$. This process is then
repeated until the \textit{full implication} $\mathcal{S}_\infty $ of $%
\mathcal{S}_0$ is determined. In $Fig.3b$ this is the set $\left\{
R,S,T,V,W,X,Y,Z\right\} .$

In this example it is assumed that each of the links is fully resolvable. If
any link is not fully resolvable, it may still be possible to construct a
non-empty full implication for certain initial data sets.

Several important concepts can be discussed with this example.

\subsection{\textbf{The future of an initial data set}}

The full implication of an initial data set consists of those events whose
event states follow \textit{from }given initial conditions, and so it is not
unreasonable to call the full implication of any initial data set the 
\textit{future} of that set. Whilst the term ``\textit{from''} in the
preceding sentence refers to a process of inference or implication carried
out in physiotime by the Theorist, the result is that diagram $Fig.3b$ for
example now carries \textit{dates }(or equivalently \textit{arrows})\textit{%
\ }as a consequence. This ordering may now be regarded as an attribute of
the mathematical model rather than of physiotime, and this is the \textit{%
mathematical arrow of time} referred to previously. The resulting structure
can then be used to represent phenomena in process time.

However, some caution should be taken here. We may encounter spacetime
dynamics which are equivalent to \textit{reversible dynamics }in continuous
time mechanics. The full implication of some initial data sets for such
spacetime mechanics may propagate both into what we would normally think of
as the conventional future \textit{and} into the conventional past. This is
in accordance with the general problem encountered with any reversible
dynamics: we can never be sure which direction is the real future and which
is the real past unless external information is supplied to tell us.

An example of a reversible discrete topology is given in $Fig.4a$. This is
the topology of the discrete time harmonic oscillator \cite
{JAROSZKIEWICZ-97A}. Assuming the links are fully resolvable, we see that
the initial data set shown (shaded and labelled $0$) has a full implication
which extends to the right and to the left of the diagram $Fig.4b$, i.e.
into the conventional past and future.

\begin{figure}[!h]
\begin{center}
\includegraphics[width=6.00in,height=3.28in]{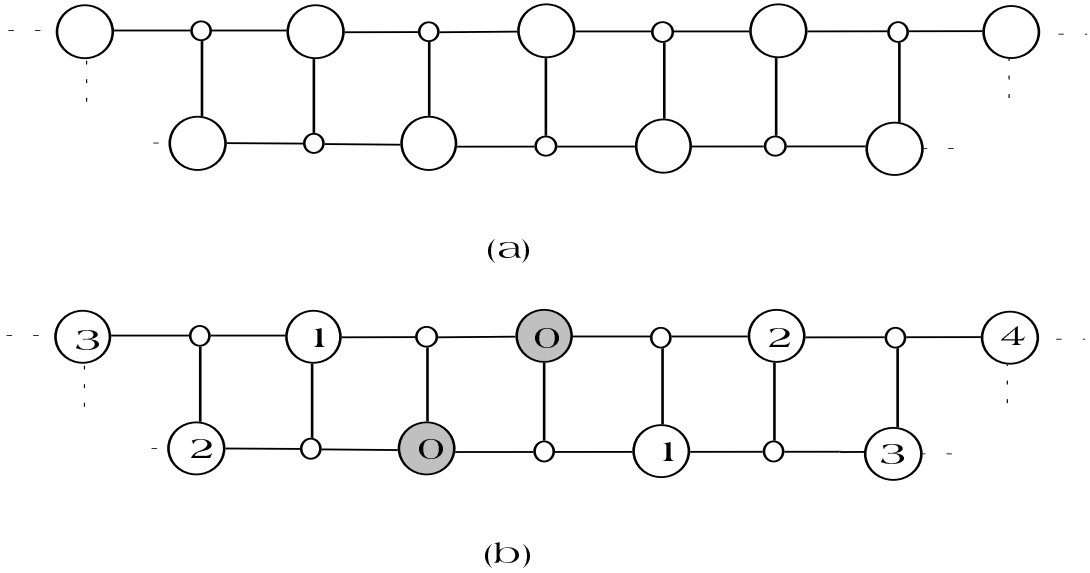}
\caption{An example of a
reversible spacetime dynamics, the discrete time harmonic oscillator. (a)
The basic discrete topology. (b) The chosen initial data set (shaded) needs
two real numbers, equivalent to initial position and velocity. The full
implication runs into the left (conventional past) and to the right
(conventional future), and there are no inaccessible events.}
\end{center}
\end{figure}

\subsection{\textbf{Inaccessible events}}

In $Fig.3b$ the set $\left\{ M,N\right\} $ is inaccessible from the given
initial data set $\left\{ O,P,Q,U\right\} $, that is, its intersection with
both the initial data set and its full implication is the empty set. Such an
inaccessible set may be interpreted in a number of ways. It could be thought
of as the \textit{absolute past }of the initial data set $\left\{
O,P,Q,U\right\} $ because in this particular example, the initial data set
implies nothing about $\left\{ M,N\right\} ,$ but specifying the states at $%
M $ and $N$ would fix $Q$. In other examples such inaccessible events could
be interpreted as beyond an event horizon of some sort. The general feature
of inaccessible events is that they cannot be affected by any changes to the
event states in an initial data set.

\textit{Archaeology }is our term for the process of reconstructing portions
of an absolute past from new and limited information added to some initial
data set. It means the process whereby specifying one or more event states
in an absolute past has the immediate consequence\textit{\ }that the
Theorist can deduce even more information about that past.

\subparagraph{\textbf{Example }$2$\textbf{: }}

Consider a spacetime dynamics based on the infinite regular triangular
topology shown in $Fig.5$, with an initial data set shaded and labelled by $%
0.$ Its first, second and third implications are labelled by $1,2$ and $3$
respectively. The full implication of the initial data set in fact extends
to infinity on the right. Suppose the Theorist now determines in some way or
decides on the state at the event shaded and dated $-1$ in $Fig.5$. Then the
events in the past of the initial data set dated by $-1$ and unshaded can
now be resolved, thereby extending the original full implication one layer to the left
of the initial data set. Just one extra piece of information can trigger an implication with an infinite number of elements. This process of retrodiction is called \textit{archaeology} for obvious reasons\footnote{%
The act of finding just a few Roman artefacts in a field may lead an Archaeologist to the conclusion that there had been a Roman settlement there.}, and
for this spacetime could be continued indefinitely into the past. The
Theorist can, by providing new initial data in this way, eventually cover the
entire spacetime as the union of an extended initial data set and its full
implication.

\begin{figure}[!h]
\begin{center}
\includegraphics[width=4.01in,height=3.89in]{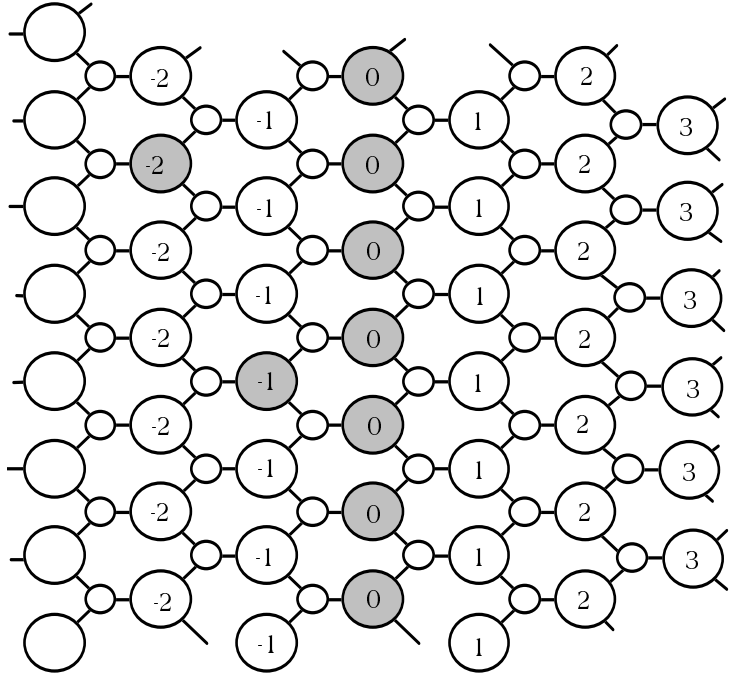}
\caption{The initial data set is
shaded and labelled by zeros. Its first, second and third implications are
labelled by \thinspace $1$,$2$, and $3$ respectively. The past of this
initial data set is the set of all events to the left of the zeros. By
providing just one extra data event (shaded and labelled $-1$) the full
implication can be extended to include the events unshaded and labelled by $%
-1$. Similarly, by providing another extra data set (shaded and labelled by $%
-2$), more evidence about what the past must have been can be deduced.}
\end{center}
\end{figure}

\section{\textbf{Causal Propagators and the speed of causality}}

Suppose we have been given an initial data set $\mathcal{S}_0$ and have
worked out its full implication $\mathcal{S}_\infty $. Now pick any event $P$
in $\mathcal{S}_0$ and change its state $\psi _P$. The consequence of this
is to change the states in some events in $\mathcal{S}_\infty $, but not
necessarily in all events in $\mathcal{S}_\infty .\;$The subset $\mathcal{P}%
_P\left( \mathcal{S}_0\right) $ of $\mathcal{S}$ consisting of all those
events changed by the change in $P$ will be called the \textit{causal
propagator }associated with\textit{\ }$P$ and $\mathcal{S}_0$. $P$ will be
called the \textit{vertex of the propagator. }

\textit{\ }We note the following:

\begin{enumerate}
\item  for any event $Q$ in an initial data set $\mathcal{S}_0,$ the causal
propagator $\mathcal{P}_Q\left( \mathcal{S}_0\right) $lies entirely within
the full implication $\mathcal{S}_\infty $ of $\mathcal{S}_0;$

\item  A causal propagator depends on a vertex \textit{and }on an associated
initial data set;

\item  A causal propagator divides spacetime into three sets: the vertex,
those events which cannot be affected by any change at the vertex, and those
events which could be changed. This structure is rather like the lightcone
structure in special relativity which separates events into those which are
timelike, lightlike, or spacelike relative to the vertex of the lightcone.
In our context, we could in some sense talk about the \textit{speed of
causality}, analogous to the speed of light in relativity, as the limiting
speed with which disturbances could propagate over our spacetime.
\end{enumerate}

\subparagraph{\textbf{Example }$3:$}

As an example we give a spacetime lattice of Type $A$ labelled by two
integers $m,n$ running from $-\infty $ to $+\infty $. The state space $%
\Omega _n^m$ at each event $P_n^m$ is the set $\left\{ +1,-1\right\} $ and a
state at $P_n^m$ will be denoted by $\psi _n^m$. The links are given by the
equations 
\[
\psi _n^m\psi _n^{m+1}\psi _n^{m-1}\psi _{n-1}^m\psi
_{n+1}^m=1,\;\;\;-\infty <m,n<\infty . 
\]
Now choose an initial data set 
\[
\psi _0^m=\psi _1^m=+1,\;\;\;-\infty <m<\infty . 
\]
This corresponds to selecting the index $m$ as a spatial coordinate and the
index $n$ as a timelike coordinate. The initial data set is then equivalent
to specifying the initial values and initial time derivatives of a scalar
field on a hyperplane of simultaneity in a two dimensional spacetime.

Because this spacetime dynamics is reversible, the full implication of this
initial data set is the entire spacetime minus the initial data set.

\begin{figure}
\begin{center}
\includegraphics[width=4.65in,height=4.71in]{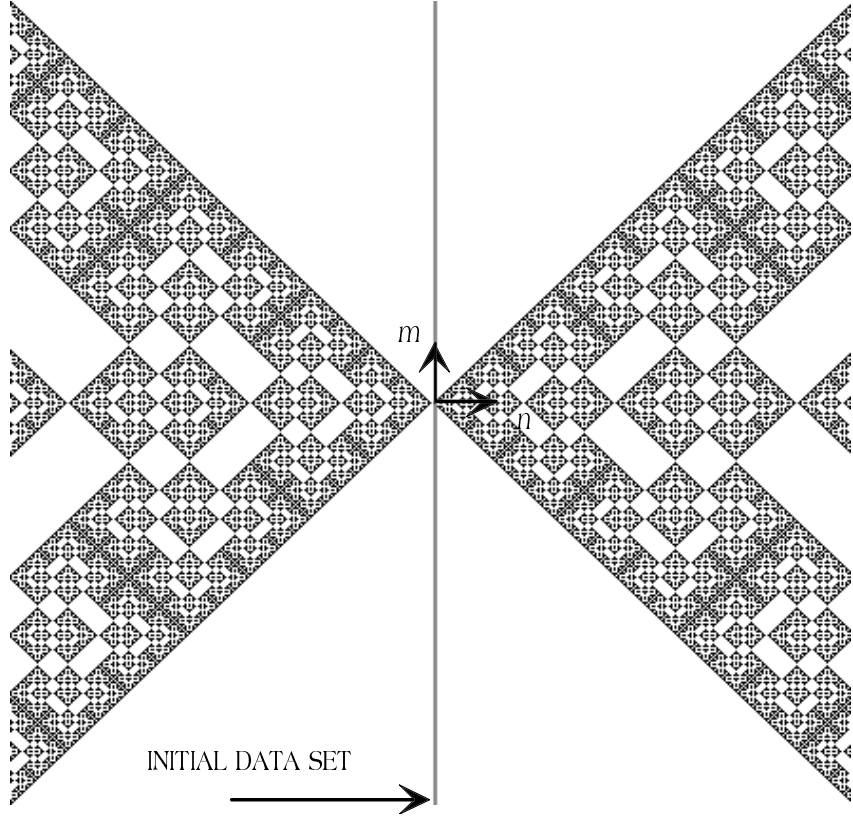}
\caption{The causal propagator
associated with the reversible spacetime dynamics and initial data set in
Example 3, with vertex at $(0,0)$. In this diagram time appears
superficially to run left to right or right to left, but actually the causal
flow is from the initial data set outwards to the left and to the right.}
\end{center}
\end{figure}

Now consider changing the state at $m=n=0$ from $\psi _0^0=+1$ to $\psi
_0^0=-1.$ In $Fig.6$ we show a bitmap plot of all those events whose state
is changed by the change in the event $\left( 0,0\right) $. The structure
looks just like a lightcone, with complex fractal-like patterns developing
inside the retarded and advanced parts of the lightcone. The speed of
causality in this example is evidently unity if we interpret the indices in
the manner discussed above.

\section{\textbf{Cosmic time}}

For some spacetime dynamics a global temporal ordering\textit{\ }can
constructed by assigning an integer to each event as follows. If $P$ is
earlier then $Q$ (i.e. \textit{P }is a partial or complete cause of $Q)$
then some integer $p$ is assigned to $P$ and some integer $q$ to $Q$ such
that $p<q.\;$These integers are called \textit{dates} above.

If it is possible to find a consistent ordering over the whole of $\mathcal{S%
}$ based on the above rule then we may say that a \textit{cosmic time}
exists for that spacetime. A cosmic time cannot be constructed for a finite
spacetime dynamics if there are no events of degree $1.$ There are two
situations where a cosmic time may be possible: either the spacetime is
finite with one or more events of degree one, or the spacetime is infinite.
However, these are not sufficient properties to guarantee a cosmic time can
be constructed.

It is possible to find spacetime dynamics for which more than one cosmic
time pattern can be established.

\subsection{\textbf{Causal (timelike) loops:}}

$Fig.7$ shows part of a spacetime dynamics containing a closed \textit{causal%
} (\textit{timelike) loop. }No cosmic time can be found for such a
spacetime. This corresponds to the situation in GR where the existence of a
closed timelike loop in a spacetime precludes the possibility of finding a
global cosmic time coordinate for that spacetime.

\begin{figure}
\begin{center}
\includegraphics[width=1.31in,height=1.42in]{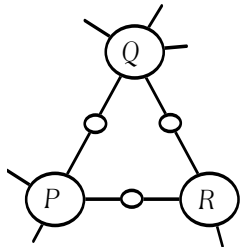}
\caption{If any spacetime
dynamics has a causal loop such as $PQR$ embedded in it then no global
cosmic time ordering is possible.}
\end{center}
\end{figure}

\subsection{\textbf{Spacelike hypersurfaces}}

In the spacetime depicted in $Fig.3$ the concept of metric was not
introduced. Nevertheless, it is possible to give a definition of a \textit{%
spacelike (hyper) }surface in this and other spacetime\textit{. }Whether
this corresponds to anything useful depends on the details of the spacetime
dynamics.

\subparagraph{\textbf{Definition 5:}}

A \textit{spacelike hypersurface }$\sum $ of a spacetime $\mathcal{S}$ is
any subset of $\mathcal{S}$ which would have a consistent full implication
if it were used as an initial data set.

We have already used the term \textit{initial data set }for such subsets.
However, not all initial data sets need be consistent. A spacelike
hypersurface is just a consistent initial data set.

There may be many spacelike hypersurfaces associated with a given spacetime
and they need not all be disjoint. The definition of spacelike hypersurface
involves a choice of causation by the Theorist. In general there may be more
than one spacelike hypersurface passing through a given event, and it is the
Theorist's choice which one to use. The possible non-uniqueness of spacelike
hypersurface associated with a given event is desirable, because this is
precisely what occurs in Minkowski spacetime, where there is an infinite
number of spacelike hypersurfaces passing through any given event,
corresponding to hyperplanes of simultaneity in different inertial frames.

\section{\textbf{Causal sets}}

Causal sets are sets with some concept of ordering relationship, which makes
them suitable for discussions concerning causality \cite{SORKIN+al-87}. This
presupposes some pre-existing temporal structure independent of any
dynamical input. This is not a feature of our dynamics, where causal
structure emerges only after the Theorist has chosen the initial data set.

\section{\textbf{Concluding remarks and summary}}

The PPM model leaves certain fundamental questions unanswered, such as the
origin of physiotime and whether process time is a meaningful concept.
However, once these are accepted as given and the model used in the right
way, then causality and time itself emerge as observer (Theorist) oriented
concepts. Many if not all of the phenomena associated with metric theories
of spacetime can be recovered, which suggests that further investigation
into this approach to spacetime dynamics may prove fruitful.

A good question to ask is: where would Lorentzian causality come from in our
approach? The answer is that it is embedded or encoded in the definition of
the links. Only when full implications are worked out from given initial
data sets would it be noticed that the dynamics itself naturally forces
certain patterns to emerge and not others. Only at that stage would the
Theorist would recognize an underlying bias in the dynamics in favour of
certain more familiar interpretations.

For example, given the continuous time equation 
\begin{equation}
\left( \frac{\partial ^2}{\partial t^2}-\frac{\partial ^2}{\partial x^2}%
-m^2\right) \varphi \left( t,x\right) =0,  \label{KG}
\end{equation}
the notation suggests that $t\,$is a time, so we could attempt to solve it
with initial data on the hyperplane $t=0.$ However, it would soon emerge
that evolution in $t$ gave runaway solutions. In other words, the equation
itself would carry the information that it would be wiser to define initial
data on the hyperplane $x=0$. We would not need to invoke the spurious
concept of Lorentzian signature metric embedded in spacetime to discover
this. At this stage we might suddenly realise that we had by some chance
interchanged the symbols for time and space in an otherwise ordinary
Klein-Gordon equation with a real mass. So rather than deal with $\left( \ref
{KG}\right) ,$ which behaves like a Klein-Gordon equation with an imaginary
mass, we would simply interchange the symbols $t$ $\,$and $x$ and then
define initial data on the hyperplane now defined by $t=0$.

\newpage\ 

\section{\textbf{Acknowledgment}}

The above is an expanded version of the author's talk at The First
International Interdisciplinary Workshop on \emph{``Studies on the Structure
of Time: from Physics to Psycho(patho)logy''}, 23-24 November 1999,
Consiglio Nazionale delle Richerche, Area della Ricerca di Palermo, Palermo,
Sicily, Italy. The proceedings of this conference, including this article,
will be published presently by Kluwer Academic (New York). 

The author is
grateful to the University of Nottingham for support, to the Organisers of
the Conference for their assistance, to the other participants for their thoughts,
and to Kluwer Academic (New York) for
permission to place this article in these electronic archives.

\end{document}